\documentclass[hidelinks,american,aps,prl,reprint,superscriptaddress]{revtex4-1}
\usepackage[latin9]{inputenc}
\usepackage{float}
\usepackage{units}
\usepackage{amsmath}
\usepackage{amssymb}
\usepackage{graphicx}
\usepackage{esint}
\usepackage{hyperref}
\hypersetup{
     colorlinks   = true,
     citecolor    = blue
}
\usepackage{color}

\makeatletter

\usepackage{braket}
\usepackage{dsfont}

\makeatother

\usepackage{babel}
\begin{document}

\title{Experimental rectification of entropy production by a Maxwell's Demon
in a quantum system}

\author{Patrice A. Camati}

\thanks{These authors contributed equally to this work.}

\affiliation{Centro de Ci\^{e}ncias Naturais e Humanas, Universidade Federal do ABC,
Avenida dos Estados 5001, 09210-580 Santo Andr\'{e}, S\~{a}o Paulo, Brazil}

\author{John P. S. Peterson}

\thanks{These authors contributed equally to this work.}

\affiliation{Centro Brasileiro de Pesquisas F\'{i}sicas, Rua Dr. Xavier Sigaud 150,
22290-180 Rio de Janeiro, Rio de Janeiro, Brazil}

\author{Tiago B. Batalh\~{a}o}

\affiliation{Centro de Ci\^{e}ncias Naturais e Humanas, Universidade Federal do ABC,
Avenida dos Estados 5001, 09210-580 Santo Andr\'{e}, S\~{a}o Paulo, Brazil}

\author{Kaonan Micadei}

\affiliation{Centro de Ci\^{e}ncias Naturais e Humanas, Universidade Federal do ABC,
Avenida dos Estados 5001, 09210-580 Santo Andr\'{e}, S\~{a}o Paulo, Brazil}

\author{Alexandre M. Souza}

\affiliation{Centro Brasileiro de Pesquisas F\'{i}sicas, Rua Dr. Xavier Sigaud 150,
22290-180 Rio de Janeiro, Rio de Janeiro, Brazil}

\author{Roberto S. Sarthour}

\affiliation{Centro Brasileiro de Pesquisas F\'{i}sicas, Rua Dr. Xavier Sigaud 150,
22290-180 Rio de Janeiro, Rio de Janeiro, Brazil}

\author{Ivan S. Oliveira}

\affiliation{Centro Brasileiro de Pesquisas F\'{i}sicas, Rua Dr. Xavier Sigaud 150,
22290-180 Rio de Janeiro, Rio de Janeiro, Brazil}

\author{Roberto M. Serra }

\email{Correspondence and requests for materials should be addressed to serra@ufabc.edu.br}

\affiliation{Centro de Ci\^{e}ncias Naturais e Humanas, Universidade Federal do ABC,
Avenida dos Estados 5001, 09210-580 Santo Andr\'{e}, S\~{a}o Paulo, Brazil}

\affiliation{Department of Physics, University of York, York YO10 5DD, United
Kingdom}
\begin{abstract}
Maxwell's demon explores the role of information in physical processes.
Employing information about microscopic degrees of freedom, this ``intelligent
observer'' is capable of compensating entropy production (or extracting
work), apparently challenging the second law of thermodynamics. In
a modern standpoint, it is regarded as a feedback control mechanism
and the limits of thermodynamics are recast incorporating information-to-energy
conversion. We derive a trade-off relation between information-theoretic
quantities empowering the design of an efficient Maxwell's demon in
a quantum system. The demon is experimentally implemented as a spin-1/2
quantum memory that acquires information, and employs it to control
the dynamics of another spin-1/2 system, through a natural interaction.
Noise and imperfections in this protocol are investigated by the assessment
of its effectiveness. This realization provides experimental evidence
that the irreversibility on a nonequilibrium dynamics can be mitigated
by assessing microscopic information and applying a feed-forward strategy
at the quantum scale.
\end{abstract}
\maketitle
Connections between thermodynamics and information theory have been
producing important insights and useful applications in the past few
years, which has turned out to be a dynamic field \cite{Kosloff2013,Goold2015,Vinjanampathy2015,Millen2016}.
Its genesis traces back to the famous Maxwell\textquoteright s demon
\textit{gedanken} experiment \cite{Leff1990,Lloyd1997,Maruyama2009,Plesch2014,Lutz2015b}.
In 1867, Maxwell conceived a \textquotedblleft neat fingered being\textquotedblright ,
which has the ability to gather information about the microscopic
state of a gas and use this information to transfer fast particles
to a hot medium and slow particles to a cold one, engendering an apparent
conflict with the second law of thermodynamics. Several approaches
and developments concerning this conundrum had been put forward \cite{Leff1990,Lloyd1997,Maruyama2009,Plesch2014,Lutz2015b},
but only after more than a century, in 1982, Bennett \cite{Bennett1982}
realized that the apparent contradiction with the second law could
be puzzled out by considering the Landauer's erasure principle \cite{Landauer1961,Landauer1988,Plenio2001,Berut2011}. 

Theoretical endeavours to incorporate information into thermodynamics
acquire a pragmatic applicability within the recent technological
progress, where information just started to be manipulated at the
micro- and nanoscale. A modern framework for these endeavors has been
provided by explicitly taking into account the change, introduced
in the statistical description of the system, due to the assessment
of its microscopic information \cite{Parrondo2015}. This outlines
an illuminating paradigm for the Maxwell\textquoteright s demon, where
the information-to-energy conversion is governed by fluctuation theorems,
which hold for small systems arbitrarily far from equilibrium \cite{Esposito2009a,Campisi2011a,Jarzynski2011,Hanggi2015,Sagawa2012}.
Generalizations of the second law in the presence of feedback control
can be obtained from this framework, establishing bounds for information-based
work extraction \cite{Sagawa2013}. Notwithstanding its fundamental
relevance, these relations do not provide a clear recipe for building
a demon in a laboratory setting. Owing to the challenges associated
with a high precision microscopic control, there are only a handful
of very recent experiments addressing the information-to-energy conversion
at small scales, using Brownian particles \cite{Toyabe2010a,Roldan2014},
single electrons \cite{koski2014a,Koski2014b-1,Koski2015}, and laser
pulses \cite{Vidrighin2016a} regarding the classical scenario, where
quantum coherence effects are absent. In the quantum context, there
are only two experimental attempts related with information-to-energy
conversion. The heat dissipated during a global system-reservoir unitary
interaction was investigated in a spin system \cite{Peterson2016}
and single photons in nonthermal states were employed to build a
thermodynamics-inspired separability criterion\textbf{ }\cite{Campini2016}.

Here, we contribute to the aforementioned efforts deriving an equality
concerning the information-to-energy conversion for a quantum nonunitary
feedback process. Such relation involves a trade-off between information-theoretic
quantities that provides a recipe to design and implement an efficient
Maxwell's demon in a quantum system where coherence is present. Supported
by this trade-off relation and employing Nuclear Magnetic Resonance
(NMR) spectroscopy \cite{Levitt2008,Oliveira2007,Jones2011}, we set
up an experimental coherent implementation of a measurement-based
feedback. Furthermore, we quantify experimentally the effectiveness
of this Maxwell's demon to rectify entropy production, due to quantum
fluctuations \cite{Batalhao2014a,Batalhao2015a}, in a nonequilibrium
dynamics. 

\begin{figure}
\includegraphics[scale=0.115]{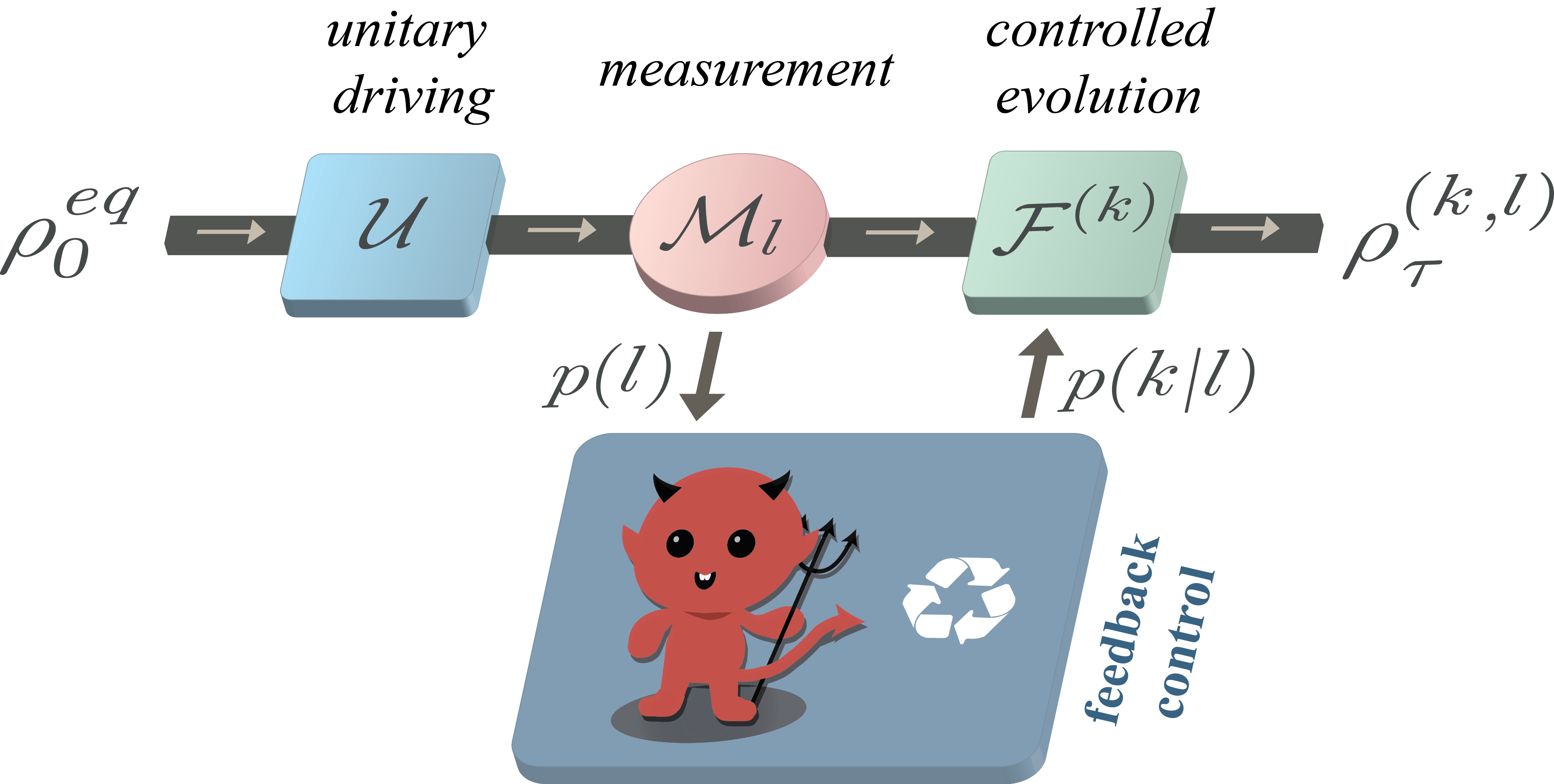}\caption{\textbf{Illustration of a Maxwell's demon operation. }The system starts
in the equilibrium state $\rho_{0}^{eq}$ and it is unitarily driven
($\mathcal{U}$) to a nonequilibrium state. Then the demon makes
a projective measurement, $\mathcal{M}$,\textbf{ }yielding the outcome
$l$ with probability $p\left(l\right)$. The feedback operation\textbf{
$\mathcal{F}^{\left(k\right)}$ }is applied with error probability
$p\left(k|l\right)$. The environment temperature is kept fixed and
the whole operation is much faster than the system decoherence time.}
\label{fig1}
\end{figure}

\emph{Theoretical description.} Consider the scenario illustrated
in Fig.~\ref{fig1}. The working system is a small quantum system,
initially in the equilibrium state $\rho_{0}^{eq}$ (at inverse temperature
\textbf{$\beta=\left(k_{B}T\right)^{-1}$}, with $k_{B}$ being the
Boltzmann constant). Later on the Maxwell's demon will also be materialized
through a microscopic quantum memory. Suppose that the working system
is driven away from equilibrium by a fast unitary time-dependent process,
$\mathcal{U}$, up to time $\tau_{1}$ (driving the system Hamiltonian
from $\mathcal{H}_{0}$ to $\mathcal{H}_{\tau_{1}}$). The purpose
of the control mechanism is to rectify the quantum fluctuations introduced
by this nonequilibrium dynamics. To this end, the demon acquires
information about the system's state through a complete projective
measurement, $\left\{ \mathcal{M}_{l}\right\} $, yielding the outcome
$l$ with probability $p\left(l\right)=\mbox{tr}\left[\mathcal{M}_{l}\mathcal{U}\rho_{0}^{eq}\mathcal{U}^{\dagger}\right]$.
Based on the outcome of this measurement a controlled evolution will
be applied. It will be described by unital quantum operations $\mathcal{F}^{\left(k\right)}$
($\mathcal{F}^{(k)}(\mathds{1})=\mathds{1}$ for every $k$), which
may include a drive on the system's Hamiltonian from $\mathcal{H}_{\tau_{1}}$
to $\mathcal{H}_{\tau_{2}}^{(k)}$, along the time interval $\tau_{2}-\tau_{1}$ \cite{Note1}.
Furthermore, we consider the possibility of error in the control mechanism,
assuming a conditional probability $p\left(k|l\right)$ of implementing
the feedback process $k$ (associated with the outcome $k$) when
$l$ is the actual observed measurement outcome. By a suitable choice
of the operations $\left\{ \mathcal{F}^{\left(k\right)}\right\} $,
the feedback control mechanism can balance out the entropy production
due to the nonequilibrium drive $\mathcal{U}$. A similar protocol
might also be employed to information-based work extraction. 

Following the scenario presented above, an integral fluctuation relation
can be derived \cite{Rastegin2013,Albash2013} as (see Supplemental Material for details):
\begin{equation}
\left\langle e^{-\beta\left(W-\Delta\text{F}^{(k)}\right)-I^{(k,l)}}\right\rangle =1,\label{eq:fluctuation relation}
\end{equation}
where \textbf{$W$} is the stochastic work done on the system, $\Delta\text{F}^{(k)}=-\beta^{-1}\ln Z_{\tau_{2}}^{(k)}/Z_{0}$
[with $Z_{t}^{(k)}=\mbox{tr}\left(e^{-\beta\mathcal{H}_{t}^{(k)}}\right)$
and $Z_{0}=\mbox{tr}\left(e^{-\beta\mathcal{H}_{0}}\right)$], is
the free energy variation for the $k$-th feedback process, $I^{(k,l)}=\ln\nicefrac{p(k|l)}{p(k)}$
is the unaveraged mutual information between the working system and
the control mechanism employed [\textbf{$p\left(k\right)=\sum_{l}p(k|l)p(l)$
}is the marginal probability distribution of the controlled operation].
The average is computed according to a work distribution probability
$P(W)$ that depends on both the measurement and the feedback processes.
Equation (\ref{eq:fluctuation relation}) has the same structure of
Sagawa and Ueda's classical relation \cite{Sagawa2010,Sagawa2012c}.
It is also the generalization of the Tasaki quantum identity obtained
for unitary control \cite{Morikuni2011a}, which was previously discussed
in Refs.~\cite{Rastegin2013,Albash2013}. Jensen's inequality for
convex functions can be used to obtain a lower bound for the mean
nonequilibrium entropy production 
\begin{equation}
\left\langle \Sigma\right\rangle \equiv\beta\left\langle W-\Delta\text{F}^{(k)}\right\rangle \geq-\left\langle I^{(k,l)}\right\rangle .\label{eq:second law like}
\end{equation}

If the feedback control is absent, Eq.~(\ref{eq:second law like})\textbf{
}reduces to the standard Clausius inequality, $\left\langle \Sigma\right\rangle \geq0$.
On the other hand, Eq.~(\ref{eq:second law like}) generalizes the
second law, elucidating that the correlations between the system and
the demon, expressed by the mutual information $\left\langle I^{(k,l)}\right\rangle $,
may be employed to decrease the entropy production beyond the conventional
thermodynamic limit. Besides its material importance to the understanding
of the underneath gear of the\textbf{ }Maxwell's demon, Eq.~(\ref{eq:second law like})
does not shed light on how to design an efficient feedback-control
protocol. Notice that the right-hand side (rhs) of Eq.~(\ref{eq:second law like})
is unrelated to the specific form of the feedback operations $\left\{ \mathcal{F}^{\left(k\right)}\right\} $,
it is only associated with the feedback error probability $p\left(k|l\right)$
and the marginal distribution $p\left(k\right)$. Therefore, performance
analysis of different types of feedback operations is beyond the scope
of the bound in Eq.~(\ref{eq:second law like}). 

We bridge such a gap by deriving an equality for entropy production
in the presence of feedback control with experimental relevance for
the effective design of a Maxwell's demon, expressed as (see Supplemental Material):
\begin{equation}
\left\langle \Sigma\right\rangle =-\mathcal{I}_{gain}+\left\langle S_{KL}\left(\rho_{\tau_{2}}^{(k,l)}||\rho_{\tau_{2}}^{(k,eq)}\right)\right\rangle +\left\langle \Delta S^{(k,l)}\right\rangle _{\mathcal{F}},\label{eq: trade-off equality}
\end{equation}
with only information-theoretic quantities on the rhs The information
gain $\mathcal{I}_{gain}=S\left(\rho_{\tau_{1}}\right)-\sum_{l}p\left(l\right)S\left(\rho_{\tau_{1}}^{\left(l\right)}\right)$
quantifies the average information that the demon obtains reading
the outcomes of the measurement $\mathcal{M}$ \cite{Groenewold1971, Lindblad1972, Ozawa1986, Fuchs2001, Buscemi2008a, Berta2014, Jacobs2014},
with $\rho_{\tau_{1}}=\mathcal{U}\rho_{0}^{eq}\mathcal{U}^{\dagger}$
being the system's state before the measurement; $\rho_{\tau_{1}}^{\left(l\right)}$
the $l$-th post-measurement state which occurs with probability $p(l)$,
and $S\left(\rho\right)$ the von Neumann entropy. It is always non-negative for projective measurements \cite{Groenewold1971, Lindblad1972, Ozawa1986, Fuchs2001} and it can be interpreted as the reciprocal to the quantity of disturbance impinged on the quantum system due to the measurement operation \cite{Buscemi2008a} (see also the Supplemental Material). The Kullback-Leibler
(KL) relative entropy, $S_{KL}\left(\rho_{\tau_{2}}^{(k,l)}||\rho_{\tau_{2}}^{(k,eq)}\right)=\mathrm{tr}\left[\rho_{\tau_{2}}^{(k,l)}\left(\ln\rho_{\tau_{2}}^{(k,l)}-\ln\rho_{\tau_{2}}^{(k,eq)}\right)\right]$,
expresses the information divergence between the resulting state of
the feedback-controlled process, $\rho_{\tau_{2}}^{(k,l)}$, and the
equilibrium state for the final Hamiltonian $\mathcal{H}_{\tau_{2}}^{(k)}$
in the $k$-th feedback process, $\rho_{\tau_{2}}^{(k,eq)}=e^{-\beta\mathcal{H}_{\tau_{2}}^{(k)}}/Z_{\tau_{2}}^{(k)}$.
The last term, $\left\langle \Delta S^{(k,l)}\right\rangle _{\mathcal{F}}=\left\langle S\left(\rho_{\tau_{2}}^{\left(k,l\right)}\right)-S\left(\rho_{\tau_{1}}^{\left(l\right)}\right)\right\rangle _{\mathcal{F}}$,
is the averaged change in von Neumann entropy due to the quantum operation
$\mathcal{F}^{(k)}$. 

The nonequilibrium entropy production in Eq.~(\ref{eq: trade-off equality})
is negative iff 
\begin{equation}
\mathcal{I}_{gain}>\left\langle S_{KL}\left(\rho_{\tau_{2}}^{(k,l)}||\rho_{\tau_{2}}^{(k,eq)}\right)\right\rangle +\left\langle \Delta S^{(k,l)}\right\rangle _{\mathcal{F}}.\label{eq: trade-off condition}
\end{equation}
This provides a necessary and sufficient condition to implement an
effective Maxwell's demon for the nonunitary protocol considered
here. Equation (\ref{eq: trade-off equality}) also encompasses the
bound $\left\langle \Sigma\right\rangle \geq-\mathcal{I}_{gain}$,
which is similar to the bounds previously obtained in Refs. \cite{Funo2013a,Sagawa2012d}
considering a different context. In the literature concerning the
thermodynamics of information, feedback processes are often regarded
as unitary. In this case the last term of the rhs of Eq.~(\ref{eq: trade-off equality})
does not contribute. Since the post-measurement state $\rho_{\tau_{1}}^{\left(l\right)}$
is pure, the average KL relative entropy, $\left\langle S_{KL}\left(\rho_{\tau_{2}}^{(k,l)}||\rho_{\tau_{2}}^{(k,eq)}\right)\right\rangle $,
will never be zero for a unitary feedback implemented upon projective
measurements (at finite temperature). In a different manner, a nonunitary
feedback process can be designed to cancel the term $\left\langle S_{KL}\left(\rho_{\tau_{2}}^{(k,l)}||\rho_{\tau_{2}}^{(k,eq)}\right)\right\rangle $,
but in this case the variation of the von Neumann entropy $\left\langle \Delta S^{(k,l)}\right\rangle _{\mathcal{F}}$,
due to a nonunitary operation, is not null. Along these lines the
trade-off concerning these quantities in Eqs. (\ref{eq: trade-off equality})
and (\ref{eq: trade-off condition}) empower the effective design
of a Maxwell's demon through the performance assessment of different
strategies for the controlled operations $\mathcal{F}^{\left(k\right)}$.

\begin{figure}
\includegraphics[scale=0.27]{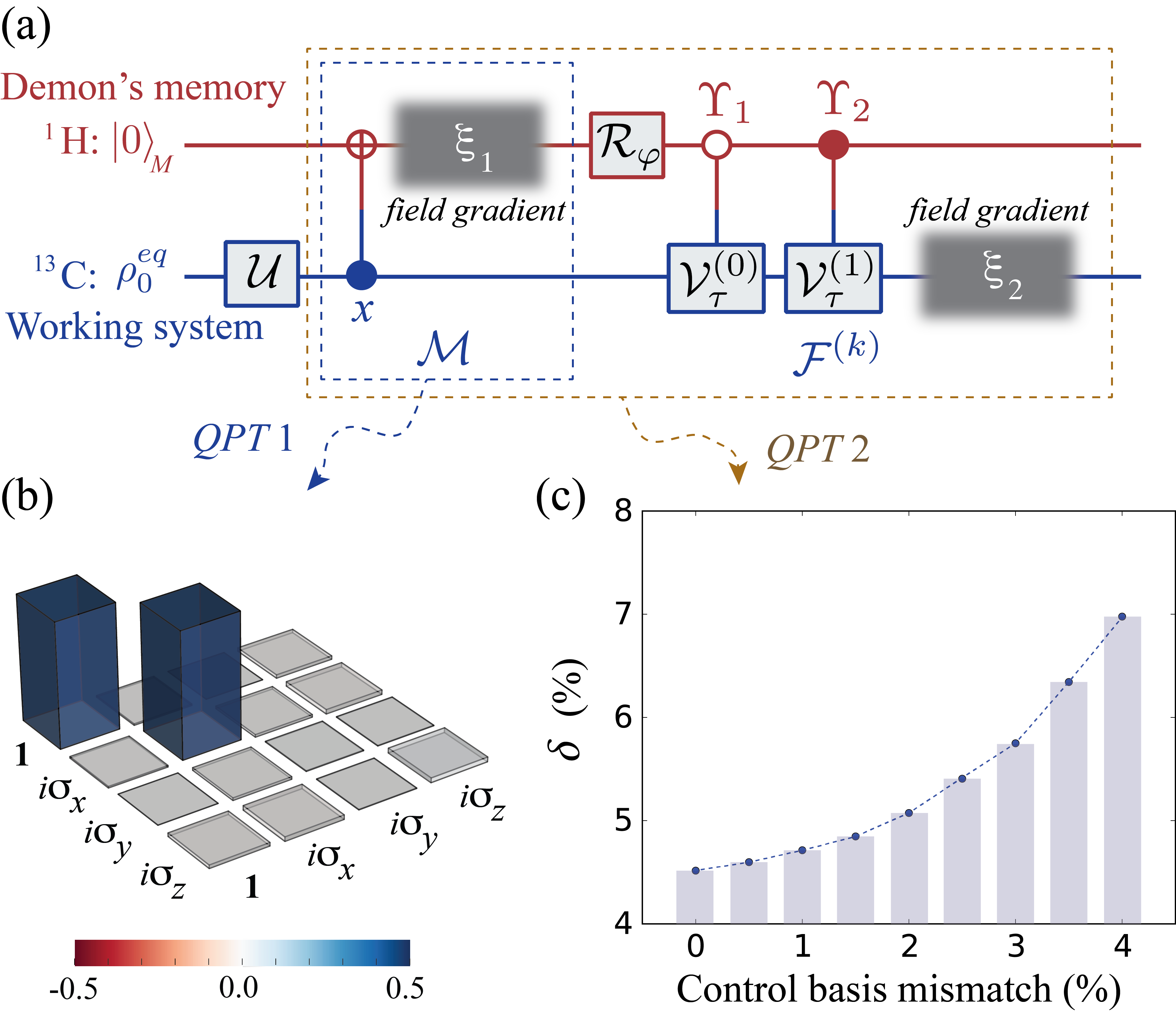}\caption{\textbf{Protocol for the measurement-based feedback. }(a) Sketch of
the implemented quantum circuit. (b) Choi-Jamiolkowski matrix, $\chi$
(with elements $\chi_{s,l}$), of the experimental quantum processes
tomography for the nonselective projective measurement on the $^{13}\text{C}$
nuclear spin, represented by the map, $\mathcal{M}\left(\rho\right)=\sum_{s,l=\mathds{1},x,y,z}\chi_{s,l}\Xi_{s}^{\text{C}}\rho\Xi_{l}^{\text{C\ensuremath{\dagger}}}$
($\Xi_{\alpha}^{\text{C}}=\mathds{1},i\sigma_{x}^{\text{C}},i\sigma_{y}^{\text{C}},i\sigma_{z}^{\text{C}}$).
The ideal process, described by a nonselective measurement on the
$\mathcal{H}_{\tau_{1}}^{\text{C}}$ energy basis, is $\mathcal{M}^{id}\left(\rho\right)=\frac{1}{2}\left(\rho+\sigma_{x}^{\text{C}}\rho\sigma_{x}^{\text{C}}\right)$.
For this operator representation choice, a unital process is described
by a real process matrix. The imaginary part of the experimental elements
$\chi_{s,l}$ are neglectful. (c) The demon effectiveness is quantified
by the process trace distance $\delta=\frac{1}{2}\text{tr}\left|\mathcal{\chi}^{exp}-\mathcal{\chi}^{id}\right|$
between the experimentally implemented map, $\mathcal{\chi}^{exp}$
(for the whole protocol: measurement and feedback control) and the
map describing the ideal protocol, $\mathcal{\chi}^{id}$, as function
of the control mismatch ($p\left(0|1\right)$). The residual error
for the zero mismatch ($\delta\approx4.5\%$) is due to nonidealities
in the protocol implementation. See main text and the Supplemental Material for details.}
\label{fig2}
\end{figure}

\emph{Experimental implementation. }We employed a $^{13}\text{C}$-labeled
$\text{CHCl}{}_{3}$ liquid sample and a 500 MHz Varian NMR spectrometer
to implement and characterize the aforementioned entropy rectification
protocol. The spin 1/2 of the $^{13}\text{C}$ nucleus is the working
system whereas the $^{1}\text{H}$ nuclear spin plays the role of
a quantum memory for the Maxwell's demon. Chlorine isotopes' nuclei
can be disregarded providing only mild environmental effects due to
the fast relaxation of its energy levels. Details on the experimental
setup are provided in the Supplemental Material. Using spatial average techniques the joint initial state,
equivalent to $\left|0\right\rangle _{\text{H}}\!\left\langle 0\right|\rho_{0}^{eq,\text{C}}$,
is prepared, where the $^{13}\text{C}$ is in an equilibrium state
of the initial Hamiltonian, $\mathcal{H}_{0}^{\text{C}}=\frac{1}{2}\hbar\omega_{0}\sigma_{z}^{\text{C}}$
(with $\frac{1}{2\pi}\omega_{0}=2$~kHz, $\sigma_{x,y,z}$ being
the Pauli matrices, $\left|0\right\rangle _{\text{H,C}}$ and $\left|1\right\rangle _{\text{H,C}}$
representing the excited and ground state of $\sigma_{z}^{\text{H,C}}$,
respectively). We consider an initial driving protocol as a sudden
quench process, described by a quick change in the Carbon Hamiltonian
from $\mathcal{H}_{0}^{\text{C}}$ to $\mathcal{H}_{\tau_{1}}^{\text{C}}=\frac{1}{2}\hbar\omega_{1}\sigma_{x}^{\text{C}}$
(with $\frac{1}{2\pi}\omega_{1}=3$~kHz). The idea is to change the
Hamiltonian so quickly that the state of the system remains unchanged.
This state will suddenly become far from equilibrium even including
coherence in the energy basis of $\mathcal{H}_{\tau_{1}}^{\text{C}}$.
The quantum fluctuations, work distribution, and the entropy production
in this highly non-adiabatic transformation can be experimentally
characterized, in an NMR setting, according the approach presented
in Refs. \cite{Batalhao2014a,Batalhao2015a}. In the present experiment,
this sudden quench is implemented effectively by a short transversal
radio-frequency (rf) pulse resonant with the $^{13}\text{C}$ nuclear
spin (with time duration about $9$ $\mu$s) represented by the operation
$\mathcal{U}$ (as in Fig.~\ref{fig2}(a)). 

\begin{figure*}
\includegraphics[scale=0.21]{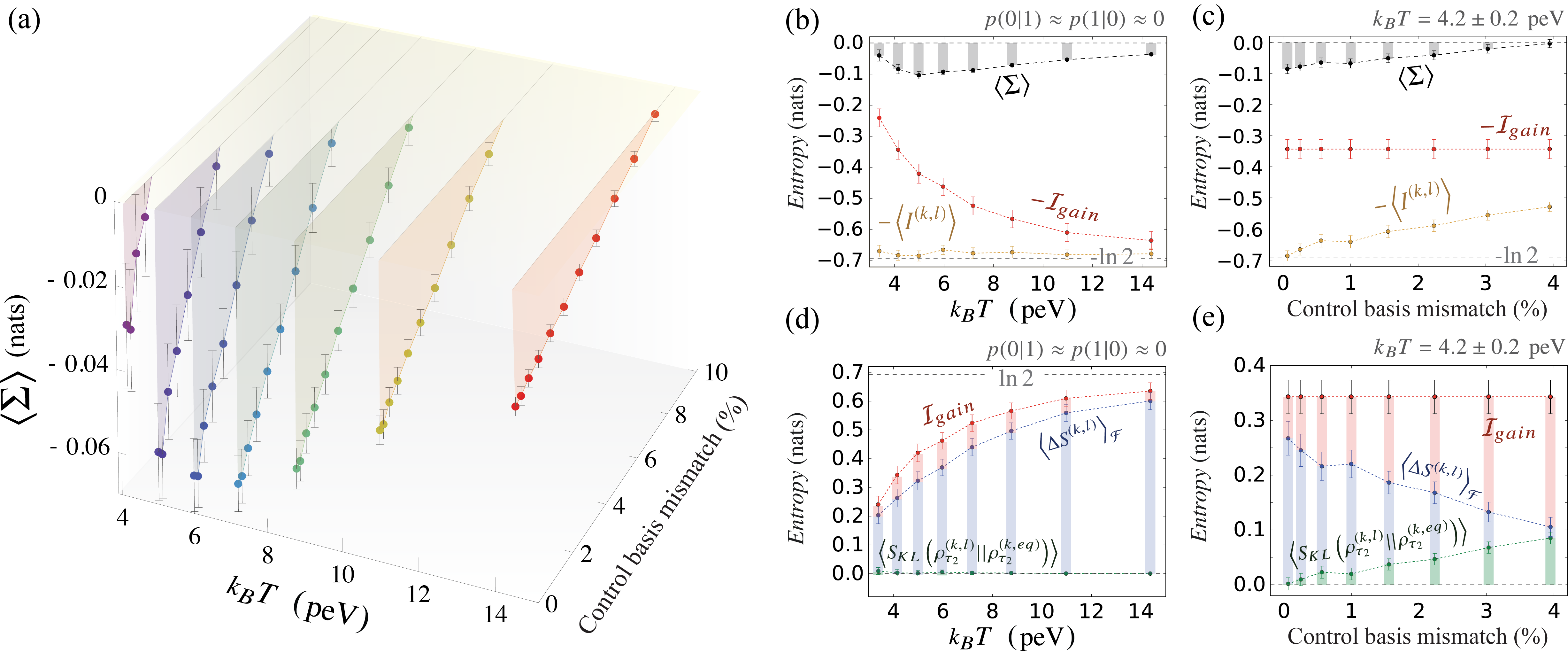}

\caption{\textbf{Experimental entropy rectification and information quantities}.
(a) Mean nonequilibrium entropy production ($\left\langle \Sigma\right\rangle $)
in the measurement-based feedback protocol as function of the initial
temperature ($k_{B}T$) and the control basis mismatch ($p(0|1)$).
The negative values are associated with entropy rectification by the
Maxwell's demon. (b) and (c) Entropy production (black), information
gain bound, $\left\langle \Sigma\right\rangle \geq-\mathcal{I}_{gain}$
(red) , and mutual information bound $\left\langle \Sigma\right\rangle \geq-\left\langle I^{(k,l)}\right\rangle $
(dark yellow). (d) and (e) Measured information quantities appearing
in the trade-off relation (\ref{eq: trade-off equality}), i.e. information
gain (red), von Neumann entropy variation (blue), and KL relative
entropy (green).
}\label{fig3}
\end{figure*}

The feedback mechanism employed is sketched in Fig.~\ref{fig2}(a), where the
whole feedback operation is much faster than the typical decoherence
times, which are on the order of seconds (see Supplemental Material). After the sudden quench ($\mathcal{U}$), information
is acquired by the demon via the natural $J$ coupling between $^{13}\text{C}$
and $^{1}\text{H}$ nuclei, $\frac{1}{2}\pi J\hbar\sigma_{z}^{\text{H}}\sigma_{z}^{\text{C}}$
(with $J=215.15$ Hz), under a free evolution lasting for about $6.97$
ms (equivalent to a CNOT gate). An effective nonselective projective
measurement in the energy basis of $\mathcal{H}_{\tau_{1}}^{\text{C}}$
is accomplished with an additional longitudinal field gradient, $\xi_{1}$
(applied during 3 ms). It introduces a full dephasing on the $z$-component
of the memory state. This free evolution followed by dehasing correlates
the state of the working system ($^{13}\text{C}$) with the demon's
memory ($^{1}\text{H}$) leading to a joint ``postmeasurement'' state
equivalent to $\left|0\right\rangle _{\text{H}}\!\left\langle 0\right|\mathcal{M}_{0}\rho_{0}^{\text{C}}\mathcal{M}_{0}+\left|1\right\rangle _{\text{H}}\!\left\langle 1\right|\mathcal{M}_{1}\rho_{0}^{\text{C}}\mathcal{M}_{1}$,
where $\mathcal{M}_{0}$ and $\mathcal{M}_{1}$ are the eigenbasis
projectors for $\mathcal{H}_{\tau_{1}}^{\text{C}}$ with experimentally
probed outcome probabilities, $p(l)=50.0\pm0.4$~\% for $l=0,1$,
as expected for the sudden quench implemented.

Quantum process tomography (QPT) \cite{Nielsen2000} is applied to
verify how effective is the demon's nonselective measurement, with
the results displayed Fig.~\ref{fig2}(b). The experimentally implemented measurement
is very close to the ideal one.\textbf{ }In order to investigate the
robustness of the feedback process against a control mismatch, we
also introduce, in the protocol of Fig.~\ref{fig2}(a), a rotation $\mathcal{R}_{\varphi}^{\text{H}}$
along the $x$-direction on the $^{1}\text{H}$ spin, in such a way
that the feedback error probability, $p\left(0|1\right)=p\left(1|0\right)=\sin^{2}\left(\frac{\varphi}{2}\right)$,
is changed varying the mismatch angle $\varphi$. Figure \ref{fig2}(c) displays
the trace distance between the experimental and ideal quantum processes
for the demon operation as a function of such an error. 

Entropy rectification is achieved by a controlled evolution of the $^{13}\text{C}$ nuclear spin
guided by the demon's memory (encoded in the $^{1}\text{H}$
nuclear spin state). Such conditional evolution is implemented by 
the operations $\Upsilon_{1}$ and $\Upsilon_{2}$
represented in Fig.~\ref{fig2}(a) produce ideally the controlled transformation
$\Upsilon_{2}\Upsilon_{1}=\left|\phi_{0}\right\rangle _{\text{H}}\!\left\langle \phi_{0}\right|\mathcal{V}_{\tau}^{(0)}+\left|\phi_{1}\right\rangle _{\text{H}}\!\left\langle \phi_{1}\right|\mathcal{V}_{\tau}^{(1)}$,
where the mismatched control basis are given by $\left|\phi_{0}\right\rangle _{\text{H}}=\cos\left(\varphi/2\right)\left|0\right\rangle _{\text{H}}-i\sin\left(\varphi/2\right)\left|1\right\rangle _{\text{H}}$
and $\left|\phi_{1}\right\rangle _{\text{H}}$ its orthogonal complement;
$\mathcal{V}_{\tau}^{(0)}=e^{-i\pi\sigma_{y}^{\text{C}}/4}e^{-i\gamma\sigma_{x}^{\text{C}}/2}$
and $\mathcal{V}_{\tau}^{(1)}=\mathcal{V}_{\tau}^{(0)}\sigma_{x}^{\text{C}}$
are the feedback operations applied on the Carbon nucleus, with $\gamma=2\,\mbox{arccos}\left(1-e^{-\beta\hbar\omega_{1}}\right)^{-1/2}$.
Both controlled operations are put into action by a free evolution
under the natural $J$ coupling ($\frac{1}{2}\pi J\hbar\sigma_{z}^{\text{H}}\sigma_{z}^{\text{C}}$)
combined with individual rotations driven by rf-fields resonant with
both Larmor frequencies of $^{13}\text{C}$ and $^{1}\text{H}$ nuclei.
We have chosen feedback operations where the system Hamiltonian is
not driven, in this case $\mathcal{H}_{\tau_{2}}^{(0)\text{C}}=\mathcal{H}_{\tau_{2}}^{(1)\text{C}}=\mathcal{H}_{\tau_{2}}^{C}=\mathcal{H}_{\tau_{1}}^{\text{C}}$.
The concluding step for implementing the controlled operations $\left\{ \mathcal{F}^{(k)}\right\} $,
is a full dephasing in the eigenbasis of $\mathcal{H}_{\tau_{2}}^{\text{C}}$.
It is supplied by a second longitudinal field gradient, $\xi_{2}$,
and local rotations of the Carbon nuclear spin in order to set the
dephasing basis. 

Performing quantum state tomography (QST) \cite{Oliveira2007} along
the experimental implementation of the demon protocol, we can obtain
all the information-theoretic quantities in rhs of Eq.~(\ref{eq: trade-off equality})
(for details see Fig. S2 and the Data Acquisition section in the Supplemental Material). 
Figure~\ref{fig3}(a) displays the entropy production in the feedback
controlled operation implemented in our experiment. We achieved negative
values showing the realization of entropy rectification, whose effectiveness
worsens as the basis mismatch increases. In Figs.~\ref{fig3}(b) and \ref{fig3}(c),
we note that the bounds based on mutual information, as in Eq.~(\ref{eq:second law like}),
and information gain are not tight in a quantum scenario, as also
anticipated by Eq.~(\ref{eq: trade-off equality}). For the present
protocol, it is possible to show that $\left\langle I^{(k,l)}\right\rangle \geq\mathcal{I}_{gain}$ (see Supplemental Material). Despite
the 4.5\% residual error in the trace distance for the zero mismatch
case [Fig.~\ref{fig2}(c)], the mutual information (between the system and
feedback mechanism) experimentally achieved is very close to its limit,
$\left\langle I^{(k,l)}\right\rangle =-\sum_{l}p\left(l\right)\ln p\left(l\right)=\ln2$~nats
(natural unit of information), as can be observed in Fig.~\ref{fig3}(b). As
discussed previously the information gain is related to how the system
correlates with the memory; hence, it is independent of the control
mismatch, which is corroborated by the experimental data in Fig.~\ref{fig3}(c).

The $k$-th feedback control operation is designed ideally to map
the Carbon spin into the equilibrium state $\rho_{\tau_{2}}^{(eq)}$
of the final Hamiltonian $\mathcal{H}_{\tau_{2}}^{\text{C}}$ (at
inverse temperature $\beta$) irrespective of the previous nonequilibrium
state $\rho_{\tau_{1}}$ (produced by the sudden quench). Our aim
is to cancel the KL relative entropy, $S_{KL}\left(\rho_{\tau_{2}}^{(k,l)}||\rho_{\tau_{2}}^{(k,eq)}\right)$,
which is successfully achieved for the zero basis mismatch, as can
be observed in Fig.~\ref{fig3}(d). On other hand the full dephasing, in the
nonunitary feedback, introduces a finite von Neumann entropy variation
$\left\langle \Delta S^{(k,l)}\right\rangle _{\mathcal{F}}$, see
Figs.~\ref{fig3}(d) and \ref{fig3}(e). This variation has no energy cost for the demon,
since it is a unital process that does not change the working system
mean energy. In the framework of the resource theory of quantum thermodynamics,
the full dephasing is regarded as a free operation \cite{Lostaglio2015,Korzekwa2016}.
When the control mismatch is increased the final state deviates from
the thermal state of $\mathcal{H}_{\tau_{2}}^{\text{C}}$ and consequently
the KL relative entropy also increases as shown in Fig.~\ref{fig3}(e). 

\emph{Discussion.} Employing an information-to-energy trade-off relation,
we designed an entropy rectification protocol based on a Maxwell's
demon. This protocol has been experimentally carried out by a coherent
implementation of a measurement-based feedback control on a quantum
spin-1/2 system. The demon's memory is a microscopic quantum ancillary
system that acquires information through a natural coupling with the
working system. Due to the quantum coherence present in our experiment,
we have to execute two dephasing operations in order to perform the
Maxwell's demon. The first dephasing operation is employed to produce
a nonselective measurement, whereas the second is essential to accomplish
entropy rectification, canceling the $\left\langle S_{KL}\left(\rho_{\tau_{2}}^{(k,l)}||\rho_{\tau_{2}}^{(k,eq)}\right)\right\rangle $
term in the trade-off relation (\ref{eq: trade-off equality}). The
present experiment elucidates the role played by different information
quantities in the quantum version of the Maxwell's demon. It also
provides evidence that the irreversibility on a quantum nonequilibrium
dynamics can be mitigated by assessing microscopic information and
applying a feed-forward strategy. The approach developed here can
be applied to general processes regarding information-to-energy conversion,
as for instance, information-based work extraction.

A future experimental challenge would be the investigation of feedback
protocols based on generalized quantum measurements and the bounds
associated with such a scheme. The analysis and the optimization of
the energetic cost for information manipulation by the Maxwell's demon,
in the quantum scenario, is also an important topic that deserves
further attention. From a broad perspective, understanding the trade-off
between information and entropy production at the quantum scale might
be important to develop applications of quantum technologies with
high efficiency. 
\begin{acknowledgments}
\emph{Acknowledgments}. We thank E. Lutz, M. Ueda, M. Herrera, and I. Henao for valuable discussions.
We acknowledge financial support from UFABC, CNPq, CAPES, FAPERJ,
and FAPESP. R.M.S. gratefully acknowledges financial support from
the Royal Society through the Newton Advanced Fellowship scheme (Grant
no. NA140436). This research was performed as part of the Brazilian
National Institute of Science and Technology for Quantum Information
(INCT-IQ).\end{acknowledgments}

\section*{Supplemental Material}
\newcommand{\beginsupplement}
{%
 \setcounter{table}{0} 
 \renewcommand{\thetable}{S\Roman{table}}%
 \setcounter{figure}{0} 
 \renewcommand{\thefigure}{S\arabic{figure}}%
 \setcounter{equation}{0} 
 \renewcommand{\theequation}{S\arabic{equation}}%
}
\beginsupplement

This supplemental material provides additional discussions and further
(theoretical and experimental) details.

\begin{figure}[h]
\includegraphics[scale=0.25]{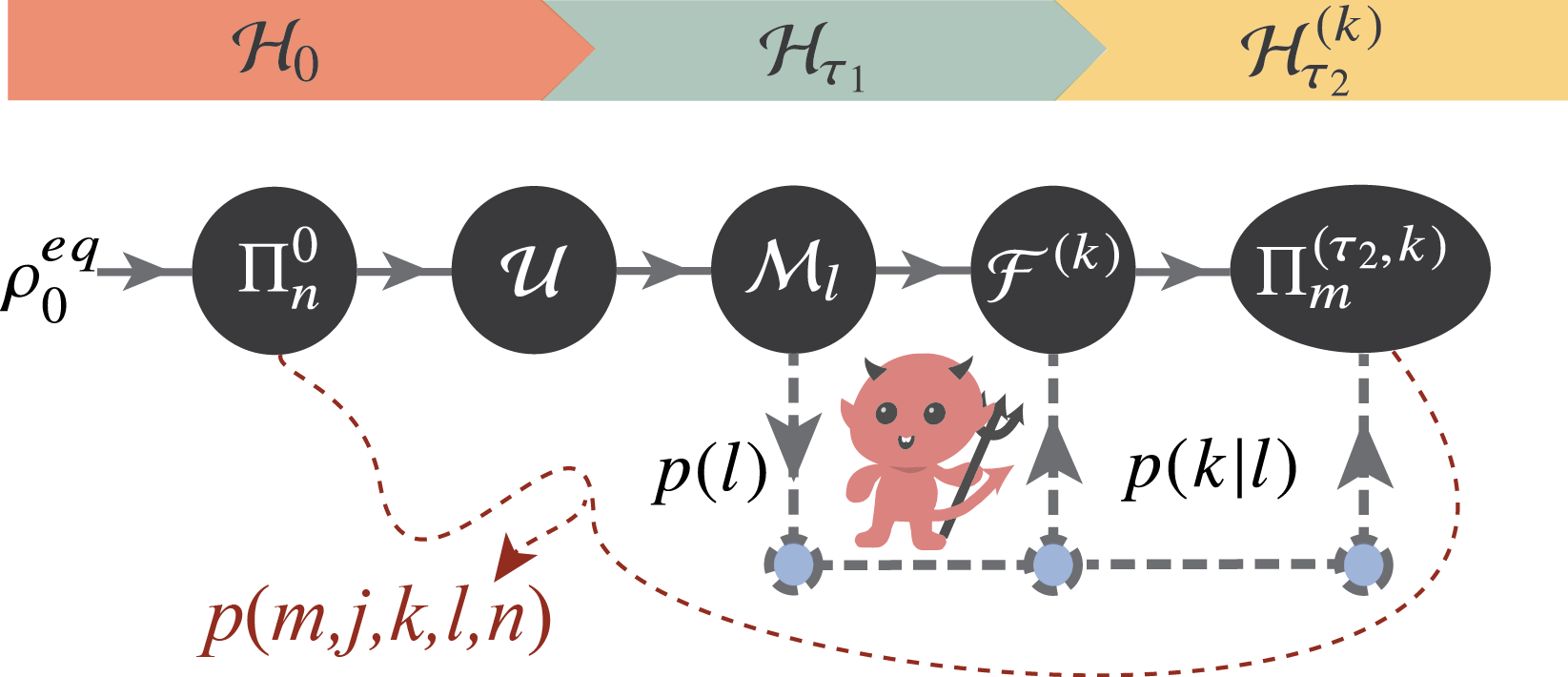}\caption{\textbf{Two-point measurement paradigm for work distribution in the
presence of feedback}. The energy basis measurements, $\Pi_{n}^{0}$,
for initial system Hamiltonian $\mathcal{H}_{0}$ and, $\Pi_{m}^{\left(\tau_{2},k\right)}$,
for $k$-th history of the feedback control (corresponding the final
Hamiltonian $\mathcal{H}_{\tau_{2}}^{\left(k\right)}$), are regarded
here as mathematical tools for the definition of the work probability
distribution. }
\label{figS1}
\end{figure}

\textbf{Work probability distribution}. In the feedback control protocol
depicted in Fig.~\ref{figS1}, the mean work done on the system is
given by the averaged work of each possible history of the feedback
process weighted by its corresponding probability
\begin{equation}
\left\langle W\right\rangle =\sum_{k,l}p\left(k,l\right)\text{U}\left(\rho_{\tau_{2}}^{\left(k,l\right)}\right)-\text{U}(\rho_{0}^{eq}),\label{eq:average work}
\end{equation}
where $p\left(k,l\right)=p\left(k|l\right)p\left(l\right)$ is the
joint probability for the $l$-th measurement outcome and $k$-th
feedback operation, $\text{U}\left(\rho_{0}^{eq}\right)=\mbox{tr}\left[\mathcal{H}_{0}\rho_{0}^{eq}\right]$
and $\text{U}\left(\rho_{\tau_{2}}^{\left(k,l\right)}\right)=\mbox{tr}\left[\mathcal{H}_{\tau_{2}}^{\left(k\right)}\mathcal{F}^{\left(k\right)}\left(\rho_{\tau_{1}}^{\left(l\right)}\right)\right]$
are the initial and final internal energy, respectively, $\rho_{\tau_{2}}^{\left(k,l\right)}=\mathcal{F}^{\left(k\right)}\left(\rho_{\tau_{1}}^{\left(l\right)}\right)$
are the possible system's final states. The operator sum decomposition
of the feedback operation is $\mathcal{F}^{\left(k\right)}\left(\cdot\right)=\sum_{j}\Gamma_{j}^{\left(k\right)}\left(\cdot\right)\Gamma_{j}^{\left(k\right)\dagger}$,
whereas the post-measurement state of the $l$-th projective measurement
is $\rho_{\tau_{1}}^{\left(l\right)}=\mathcal{M}_{l}\mathcal{U}\rho_{0}^{eq}\mathcal{U}^{\dagger}\mathcal{M}_{l}/p\left(l\right)$.
Since the unital processes considered here do not involve energy exchange
with the reservoir, the change in the internal energy is regarded
as work. Using the spectral decomposition of both Hamiltonians, $\mathcal{H}_{0}=\sum_{n}\varepsilon_{n}^{(0)}\Pi_{n}^{0}$
and $\mathcal{H}_{\tau_{2}}^{\left(k\right)}=\sum_{m}\varepsilon_{m}^{\left(\tau_{2},k\right)}\Pi_{m}^{\left(\tau_{2},k\right)}$,
one can write Eq.~(\ref{eq:average work}) as

\begin{eqnarray}
\left\langle W\right\rangle  & = & \sum_{m,j,k,l,n}p\left(k|l\right)p\left(m,j,l,n\right)\Delta\varepsilon_{m,n}^{\left(k\right)}\nonumber \\
 & = & \sum_{m,j,k,l,n}p\left(m,j,k,l,n\right)\Delta\varepsilon_{m,n}^{\left(k\right)},\label{eq:average}
\end{eqnarray}
with $p\left(m,j,l,n\right)\equiv\mbox{tr}\left(\Pi_{m}^{\left(\tau_{2},k\right)}\Gamma_{j}^{\left(k\right)}\mathcal{M}_{l}\mathcal{U}\Pi_{n}^{(0)}\rho_{0}^{eq}\mathcal{U}^{\dagger}\mathcal{M}_{l}\Gamma_{j}^{\left(k\right)\dagger}\right)$,
$p\left(m,j,k,l,n\right)\equiv p\left(k|l\right)p\left(m,j,l,n\right)$,
and $\Delta\varepsilon_{m,n}^{\left(k\right)}=\varepsilon_{m}^{\left(\tau_{2},k\right)}-\varepsilon_{n}^{(0)}$.
We can express the work distribution in the presence of feedback as
$P\left(W\right)=\sum_{m,j,k,l,n}p\left(m,j,k,l,n\right)\delta\left(W-\Delta\varepsilon_{m,n}^{\left(k\right)}\right)$
and the average work as $\left\langle W\right\rangle =\int dW\,P\left(W\right)W$.
This work distribution can also be related to the two-point energy
measurement paradigm~\cite{Talkner2007}, considering a measurement
on the energy basis of $\mathcal{H}_{0}$ at the beginning of the
protocol described in Fig.~\ref{figS1} (before the unitary driven
$\mathcal{U}$) and another measurement at the end of the protocol
in the energy basis of $\mathcal{H}_{\tau_{2}}^{\left(k\right)}$
in the $k$-th history. Notice that in the presence of the feedback,
the second measurement of the two-point paradigm depends on the feedback
operation implemented since it can drive the Hamiltonian to $\mathcal{H}_{\tau_{2}}^{\left(k\right)}$
as illustrated in Fig.~\ref{figS1}.

\textbf{Fluctuation relation in the presence of a unital feedback}.
For the sake of completeness, we will verify the validity of the fluctuation
relation in Eq.~(\ref{eq:fluctuation relation}) of the main text, which was previously discussed
in Refs.~\cite{Rastegin2013,Albash2013}. Consider the following
average
\begin{align}
\left\langle e^{-\beta\left(W-\Delta\text{F}^{\left(k\right)}\right)-I^{\left(k,l\right)}}\right\rangle  & =\sum_{m,j,k,l,n}p\left(m,j,k,l,n\right)\nonumber \\
 & \times e^{-\beta\left(\Delta\varepsilon_{m,n}^{\left(k\right)}-\Delta F^{\left(k\right)}\right)-I^{\left(k,l\right)}.}\nonumber \\
 & =\sum_{m,j,k,l,n}p\left(m,j,l,n\right)\nonumber \\
 & \times Z_{0}e^{+\beta\varepsilon_{n}^{(0)}}\frac{e^{-\beta\varepsilon_{m}^{\left(\tau_{2},k\right)}}}{Z_{\tau_{2}}^{\left(k\right)}}p\left(k\right).\label{eq:aux2}
\end{align}

Remembering the definition of $p\left(m,j,l,n\right)$ introduced
in the previous section and identifying $\Pi_{n}^{(0)}\rho_{0}^{eq}=Z_{0}^{-1}e^{-\beta\varepsilon_{n}^{0}}\Pi_{n}^{(0)}$
and $\rho_{\tau_{2}}^{\left(k,eq\right)}=\frac{1}{Z_{\tau_{2}}^{\left(k\right)}}\sum_{m}e^{-\beta\varepsilon_{m}^{\left(\tau_{2},k\right)}}\Pi_{m}^{\left(\tau_{2},k\right)}$;
the right hand side (rhs) of Eq.~(\ref{eq:aux2}) can be simplified
to $\sum_{k,j,l}p\left(k\right)\,\mbox{tr}\left(\rho_{\tau_{2}}^{\left(k,eq\right)}\Gamma_{j}^{\left(k\right)}\mathcal{M}_{l}\Gamma_{j}^{\left(k\right)\dagger}\right)$.
Using the completeness of the measurement and the unitality of the
map, $\mathcal{F}^{(k)}(\mathds{1})=\sum_{k}\Gamma_{j}^{\left(k\right)}\Gamma_{j}^{\left(k\right)\dagger}=\mathds{1}$,
Eq.~(\ref{eq:aux2}) turns out to be 
\[
\left\langle e^{-\beta\left(W-\Delta\text{F}^{\left(k\right)}\right)-I^{\left(k,l\right)}}\right\rangle =\sum_{k}p\left(k\right)\,\mbox{tr}\left(\rho_{\tau_{2}}^{\left(k,eq\right)}\right).
\]
Since $\mbox{tr}\left(\rho_{\tau_{2}}^{\left(k,eq\right)}\right)=1$
and $p\left(k\right)$ is a normalized distribution, we obtain the
fluctuation relation in Eq.~(\ref{eq:fluctuation relation}) of the main text.

\textbf{Derivation of the trade-off relation}. Let us start from KL
relative entropy between an arbitrary state, $\rho$, and the equilibrium
state, $\rho^{eq}$, associated with the Hamiltonian $\mathcal{H}$
at inverse temperature $\beta$, 
\begin{eqnarray}
S_{KL}\left(\rho||\rho^{eq}\right) & = & \mathrm{tr}\left[\rho\left(\ln\rho-\ln\rho^{eq}\right)\right]\nonumber \\
 & = & \mathrm{-tr}\left(\rho\ln\frac{e^{-\beta\mathcal{H}}}{Z}\right)-S\left(\rho\right)\nonumber \\
 & = & \mathrm{\beta tr}\left(\rho\mathcal{H}\right)-\ln Z-S\left(\rho\right)\nonumber \\
 & = & \beta\left[\text{U\ensuremath{\left(\rho\right)}}-\text{F}\right]-S\left(\rho\right).\label{eq:identityutil}
\end{eqnarray}
From the above identity we can write $\beta U\left(\rho_{\tau_{2}}^{\left(k,l\right)}\right)=S_{KL}\left(\rho_{\tau_{2}}^{\left(k,l\right)}||\rho_{\tau_{2}}^{\left(k,eq\right)}\right)+\beta\text{F}_{\tau_{2}}^{(k)}+S\left(\rho_{\tau_{2}}^{\left(k,l\right)}\right)$
and $\beta U\left(\rho_{0}^{eq}\right)=\beta\text{F}_{0}+S\left(\rho_{0}^{eq}\right)$,
with $\text{F}_{\tau_{2}}^{(k)}=-\beta^{-1}\ln Z_{\tau_{2}}^{(k)}$
and $\text{F}_{0}=-\beta^{-1}\ln Z_{0}$ . These results combined
with Eq.~(\ref{eq:average work}) lead to the following expression
for the mean nonequilibrium entropy production in the presence of
feedback: 
\begin{eqnarray}
\left\langle \Sigma\right\rangle  & = & \beta\left\langle W-\Delta\text{F}^{(k)}\right\rangle \nonumber \\
 & = & -S\left(\rho_{0}^{eq}\right)+\sum_{l}p\left(l\right)S\left(\rho_{\tau_{1}}^{\left(l\right)}\right)\nonumber \\
 &  & +\sum_{k,l}p\left(k,l\right)S_{KL}\left(\rho_{\tau_{2}}^{\left(k,l\right)}||\rho_{\tau_{2}}^{\left(k,eq\right)}\right)\nonumber \\
 &  & +\sum_{k,l}p\left(k,l\right)\left(S\left(\rho_{\tau_{2}}^{\left(k,l\right)}\right)-S\left(\rho_{\tau_{1}}^{\left(l\right)}\right)\right),\label{eq:auxfinal}
\end{eqnarray}
where we have added and subtracted the averaged entropy $\sum_{l}p\left(l\right)S\left(\rho_{\tau_{1}}^{\left(l\right)}\right)$.
The rhs of Eq.~(\ref{eq:auxfinal}) can be identified with the
information-theoretic quantities (i.e. information gain, KL relative
entropy, and the von Neumann entropy variation, respectively) resulting
in the trade-off relation in Eq. ~(\ref{eq: trade-off equality}) of the main text.

\textbf{Information gain.} When performing a measurement on a quantum system, the observer acquires
information about the system. The description of such a process plays
a central role in quantum measurement theory \cite{Jacobs2014}. The
information gain was first proposed by Groenewold in 1971 \cite{Groenewold1971}
whose aim was to quantify how much information is obtained by a quantum
measurement. Groenewold proposed that the information gained by the
observer is given by the average uncertainty reduction of the quantum
state, and its given by
\begin{equation}
\mathcal{I}_{gain}=S\left(\rho\right)-\sum_{l}p\left(l\right)S\left(\rho^{(l)}\right),\label{eq:igain-sup}
\end{equation}
where $\rho$ is the state immediately before the measurement and
$\rho^{(l)}$ the post-measurement states, which occur with outcome
probability $p(l)$. A year after its introduction, Lindblad showed
that the information gain, as expressed in Eq.~(\ref{eq:igain-sup}),
is non-negative for von Neumann projective measurements \cite{Lindblad1972}.
Over a decade later, Ozawa generalized the demonstration of Lindblad
showing that the information gain is non-negative for any positive-operator
valued measure (POVM) \cite{Ozawa1986}. POVMs and von Neumann measurements
are called efficient measurements because each measurement operator
is associated with a single outcome (each measurement operator is
described by a single Kraus operator). On the other hand, measurements
which possess an intrinsic classical uncertainty are called inefficient.
Inefficient measurements induce more back-action to the system than
quantum mechanics would allow for the same amount of information extracted
\cite{Jacobs2014}. Equation~(\ref{eq:igain-sup}) may becomes negative
for such inefficient measurements and therefore may not be a good
quantifier in this particular case. Recently, Buscemi, Hayashi,
and Horodeck \cite{Buscemi2008a} generalized the expression
for the information gain providing a new framework which gives positive
results for any kind of measurement (efficient or inefficient) and
recovers Eq.~(\ref{eq:igain-sup}) for efficient measurements. Furthermore,
they endorsed the interpretation of the information gain by providing
an operational meaning to it. In order to approach this very general
class of measurements, in Ref.~\cite{Buscemi2008a} the measuring apparatus
is composed by three ancillary systems apart from the system to be
measured (so-called quantum instruments). One ancilla is employed
as a purification system and the other two to emulate the desired
measurement, similar to the role of the quantum memory in the measurement
stage of our feedback protocol. This newly proposed information gain
is based on the mutual information of these auxiliary systems, although
ultimately it depends only on quantities of the system. 

In the present article, we have considered projective measurements
(which are experimentally implemented with a very high accuracy assessed
by quantum tomography), therefore the Groenewold definition of information
gain, in Eq.~(\ref{eq:igain-sup}), is suitable to be applied throughout
all of our analyzes. We also note that Eq.~(\ref{eq: trade-off equality}) of the main text holds
for a general projective measurement feedback control based on unital
operations. An interesting point for future investigation is the possibility
to derive an information thermodynamics trade-off relation for a feedback
mechanism involving inefficient measurements. 

For the sake of clarity, we provide below a short demonstration
(based on Refs. \cite{Jacobs2014,Fuchs2001}) that the information
gain as defined in Eq.~(\ref{eq:igain-sup}) is non-negative for POVM
measurements. Any operator $A$ may be decomposed as $A=PU$,
where $P$ is a positive operator and $U$ is a unitary operator.
This decomposition is called the polar decomposition of $A$. Using
this decomposition we can show that $AA^{\dagger}=P^{2}$. Therefore,
$A^{\dagger}A=U^{\dagger}P^{2}U=U^{\dagger}AA^{\dagger}U$, which
means that the operators $A^{\dagger}A$ and $AA^{\dagger}$ possess
the same set of eigenvalues (since unitary transformation preserve
eigenvalues). For a POVM, $\left\{ M_{l}^{\dagger}M_{l}\right\} $,
the $l$-th post-measurement state, $\rho^{(l)}=M_{l}\rho M_{l}^{\dagger}/p(l)$,
is obtained with probability $p(l)=\mbox{tr}\left(M_{l}^{\dagger}M_{l}\rho\right)$.
The measurement operators of the POVM satisfy the completeness relation $\sum_{l}M_{l}^{\dagger}M_{l}=\mathds{1}$.
We can write the state before the measurement, $\rho,$ in the following
convenient form
\begin{equation}
\rho=\sqrt{\rho}\mathds{1}\sqrt{\rho}=\sum_{l}\sqrt{\rho}M_{l}^{\dagger}M_{l}\sqrt{\rho}=\sum_{l}p(l)\mu^{(l)},
\end{equation}
where $\mu^{(l)}=\sqrt{\rho}M_{l}^{\dagger}M_{l}\sqrt{\rho}/p(l)$.
The von Neumann entropy is a concave function, which means $S\left(\sum_{n}p(n)\rho^{(n)}\right)\geq\sum_{n}p(n)S\left(\rho^{(n)}\right)$
with $\sum_{n}p(n)=1$. From the concavity of the von Neumann entropy
we obtain
\begin{equation}
S\left(\rho\right)\geq\sum_{l}p(l)S\left(\mu^{(l)}\right).
\end{equation}
Defining $X_{l}=\sqrt{\rho}M_{l}/\sqrt{p(l)}$ we write $\mu^{(l)}=X_{l}X_{l}^{\dagger}$.
So $\mu^{(l)}$ has the same eigenvalues as $\rho^{(l)}=X_{l}^{\dagger}X_{l}=M_{l}\rho M_{l}^{\dagger}/p(l)$.
Since the von Neumann entropy is a function of only the eigenvalues
of the density operator $S\left(\mu^{(l)}\right)=S\left(\rho^{(l)}\right)$,
which implies the non-negativity of the information gain defined in
Eq.~(\ref{eq:igain-sup}) for any POVM.

\textbf{Experimental set-up.} The liquid sample consist of $50$~mg
of $99$\% $^{13}\text{C}$-labeled $\text{CHCl}_{3}$ (Chloroform)
diluted in $0.7$ ml of $99.9$\% deutered Acetone-d6, in a flame
sealed Wildmad LabGlass 5 mm tube. All experiments are carried out
in a Varian $500$~MHz Spectrometer employing a double-resonance
probe-head equipped with a magnetic field gradient coil. Chloroform
sample is very diluted so that the intermolecular interaction can
be neglected, and the sample can be regarded as a set of identically
prepared pairs of spin-1/2 systems. The sample is placed in the presence
of a longitudinal static magnetic field (whose direction is taken
to be along the positive $z$ axes) with strong intensity, $B_{0}\approx11.75$
T. The nuclear magnetization of $^{1}\text{H}$ and $^{13}\text{C}$
precess around $B_{0}$ with Larmor frequencies about $500$~MHz
and $125$~MHz, respectively. Magnetization of the nuclear spins
are controlled by time-modulated rf-field pulses in the transverse
($x$ and $y$) direction and longitudinal field gradients. 

Spin-lattice relaxation times, measured by the inversion recovery
pulse sequence, are \foreignlanguage{english}{$\left(\mathcal{T}_{1}^{H},\mathcal{T}_{1}^{C}\right)=\left(7.42,\:11.31\right)$~}s.
Transverse relaxations, obtained by the Carr-Purcell-Meiboom-Gill
(CPMG) pulse sequence, have characteristic times \foreignlanguage{english}{$\left(\mathcal{T}_{2}^{\text{*H}},\mathcal{T}_{2}^{*\text{C}}\right)=\left(1.11,\:0.30\right)$}~s.
The total experimental running time, to implement the entropy rectification
protocol, is about $22.4$\textbf{\footnotesize{}~}ms, which is considerably
smaller than the spin-lattice relaxation and therefore decoherence
can be disregard. The data for the process tomography, showed in Fig.~\ref{fig2}
of the main text, also endorses this consideration, since the experimentally
implemented process does not exhibit significant decoherence effects. 

The initial state of the nuclear spins is prepared by spatial average
techniques \cite{Oliveira2007,Jones2011,Batalhao2014a,Batalhao2015a},
being $^{1}\text{H}$ nucleus prepared in the ground state and the
$^{13}\text{C}$ nucleus in a pseudo-thermal state with the populations
(in the energy basis of $\mathcal{H}_{0}^{\text{C}}$) and corresponding
pseudo-temperatures displayed in Tab.~\ref{tabSI}.

\begin{table}[H]
\caption{\textbf{Population and pseudo-temperature of the Carbon initial states. }}
{
\begin{center}
\footnotesize
$\begin{array}{cccc}
\hline
p_0^{(0)} & p_1^{(0)}  & k_{B}T \, \text{(peV)}\\
\hline
\hline
0.96\pm0.01 & 0.04\pm0.01 & 2.6\pm0.2\\
0.92\pm0.01 & 0.08\pm0.01 & 3.4\pm0.2\\
0.88\pm0.01 & 0.12\pm0.01 & 4.2\pm0.2\\
0.84\pm0.01 & 0.16\pm0.01 & 4.9\pm0.2\\
0.81\pm0.01 & 0.19\pm0.01 & 5.9\pm0.3\\
0.76\pm0.01 & 0.24\pm0.01 & 7.0\pm0.3\\
0.73\pm0.01 & 0.27\pm0.01 & 8.6\pm0.4\\
0.69\pm0.01 & 0.31\pm0.01 & 10.7\pm0.6\\
0.65\pm0.01 & 0.35\pm0.01 & 13.8\pm1.0\\
\hline
\end{array}$
\label{tabSI}
\end{center}
}

\end{table}

\begin{figure}[h]
\includegraphics[scale=0.22]{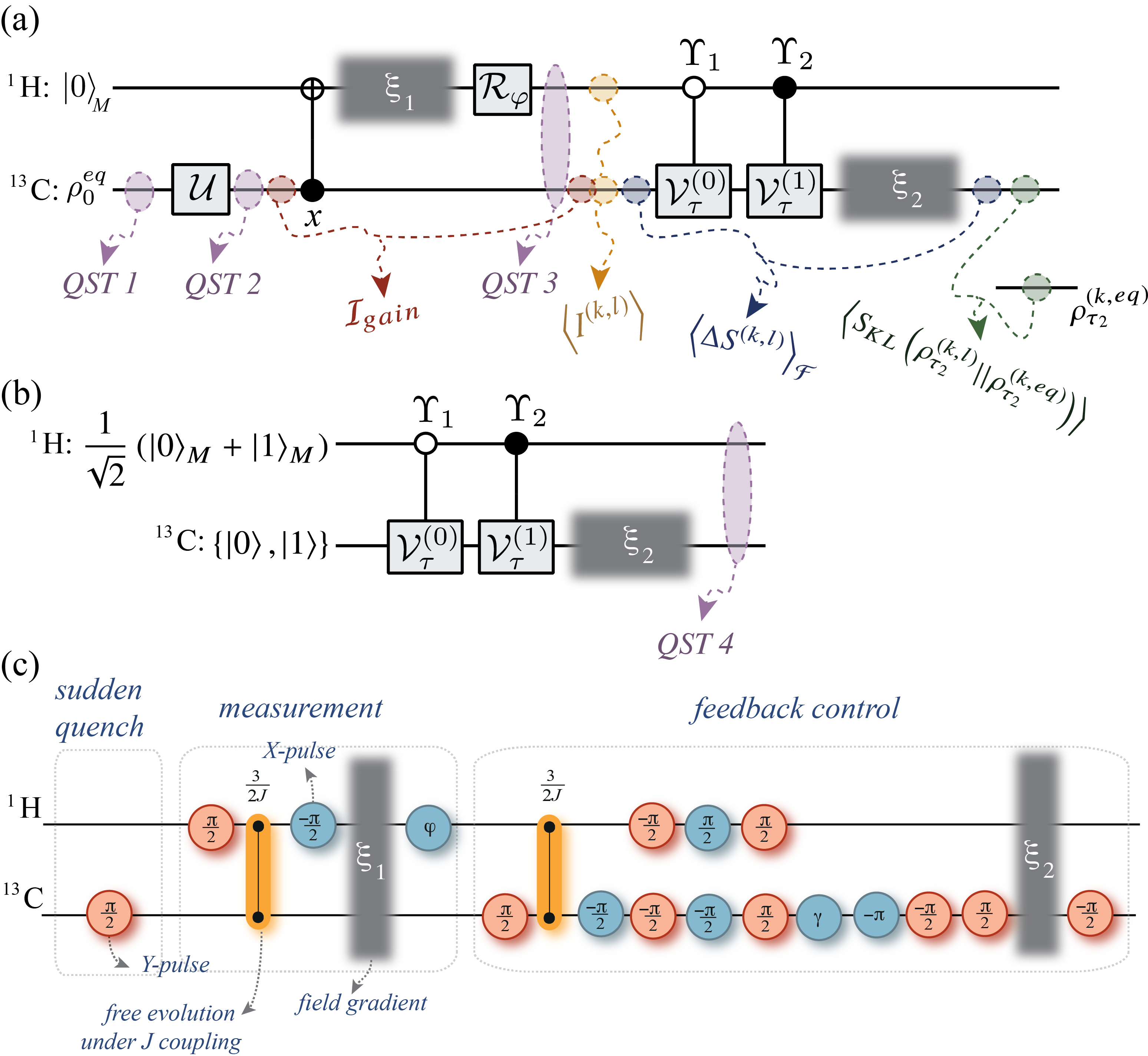}\caption{\textbf{Characterization of the Maxwell's demon operation protocol.
}(a) and (b) Quantum state tomography strategy to obtain the relevant
information quantities. The initial states, as displayed, are prepared
and the feedback control is applied. This enables the full information
characterization of the feedback controlled evolution. QST~1 and
QST~2 are single spin-1/2 tomography realized on the Carbon nucleus,
whereas QST~3 and QST~4 are joint tomographies implemented in both
Hydrogen and Carbon nuclei. (c) Optimized pulse sequence used to implement
the measurement-based feedback control operation. Blue (red) circles
represent one spin transverse rf-pulses producing rotations on the
$x$ ($y$) by the displayed angle. Free evolutions under the natural
$J$ coupling ($\frac{1}{2}\pi J\hbar\sigma_{z}^{\text{H}}\sigma_{z}^{\text{C}}$)
are represented by two-spins connections (in orange), with the time-length
displayed above the connections. The grey regions represent the two
longitudinal field gradients.}
\label{figS2}
\end{figure}

\textbf{Data acquisition.} Quantum state tomography (QST) is employed to
obtain the relevant information quantities in the controlled feedback
process. We have performed QST along the protocol implementation 
of the demon and we have employed an auxiliary circuit as depicted in 
Figs.~\ref{figS2}(a) and \ref{figS2}(b), respectively. QST~1 is
used to verify the effective temperature of the initial state. From
QST~2 and QST~3 we obtain the information gain, $\mathcal{I}_{gain}$.
The mutual information, $\left\langle I^{(k,l)}\right\rangle $ is
obtained from QST~3. The remaining information quantities ($\left\langle S_{KL}\left(\rho_{\tau_{2}}^{(k,l)}||\rho_{\tau_{2}}^{(k,eq)}\right)\right\rangle $
and $\left\langle \Delta S^{(k,l)}\right\rangle _{\mathcal{F}}$)
are obtained from the aforementioned tomographic data combined with
QST~4 (obtained from the auxiliary circuit in Fig.~\ref{figS2}(b)). Due to the fact that our implementation is based on nonselective measurements, an auxiliary circuit (Fig.~\ref{figS2}(b)) is employed to obtain the states  $\rho_{\tau_{2}}^{(k,l)}$ by QST. This enables us to characterize the information quantities in rhs of Eq.~(\ref{eq: trade-off equality}) in the main text. The optimized pulse sequence used to implement the Maxwell's demon is displayed in Fig.~\ref{figS2}(c).

The quantum process tomography, as illustrated in Fig.~\ref{fig2}(a) of the
main text, is carried out by preparing a set of mutually unbiased
basis (MUB) states \cite{Nielsen2000}, implementing the nonselective
projective measurement operation (the full controlled feedback protocol)
in QPT~1 (in QPT~2) and then a full quantum state tomography at
the end. From this data it is possible to obtain the Choi-Jamiolkowski
matrix, $\chi$, of the process. The error in the Maxwell's demon
realization is probed by the process trace distance, $\delta=\frac{1}{2}\text{tr}\left|\mathcal{\chi}^{exp}-\mathcal{\chi}^{id}\right|$,
between the ideal ($id$) and the experimentally ($exp$) implemented
processes, as displayed in Fig.~\ref{fig2}(c) of the main text. Its operational
interpretation is related with the bias for the distinguishability
between the ideal and experimental processes. The average success
probability when distinguishing the two processes is $\frac{1}{2}+\frac{1}{2}\delta$,
when both processes are performed with equal a priori probability.

\textbf{Information bounds for entropy production. }For the projective
nonselective measurement implemented in our protocol, the information
gain reduces to $\mathcal{I}_{gain}=S\left(\rho_{\tau_{1}}\right)$,
in the ideal case. Since the driving process implemented, $\mathcal{U}$,
is a unitary sudden quench, the von Neumann entropy after the quench,
$S\left(\rho_{\tau_{1}}\right)$, is the same as for the initial equilibrium
state, $S\left(\rho_{0}^{eq}\right)$. This latter entropy is also
equal to the classical Shannon entropy $\textsf{H}_{Sh}(p_{0}^{(0)},p_{1}^{(0)})=-\sum_{i=0,1}p_{i}^{(0)}\ln p_{i}^{(0)}$,
of the populations in the initial Hamiltonian ($\mathcal{H}_{0}$)
energy basis (with $p_{i}^{(0)}=\text{tr}\left(\Pi_{i}\rho_{0}^{eq}\right)$).
So $\mathcal{I}_{gain}=\textsf{H}_{Sh}(p_{0}^{(0)},p_{1}^{(0)})$.
On the other hand, the probability for the $l$-th measurement outcome
after the sudden quench is equally weighted in the ideal case, $p(l)=\frac{1}{2}$,
for $l=0,1$. In the absence of basis mismatch, the correlation generated,
by the coherent implementation of the measurement-based feedback,
leads ideally to the joint probability distribution [$p(k,l)$] of
the controlled operation ($k$) and measurement outcome ($l$), $p(0,0)=p(1,1)=\frac{1}{2}$
and $p(0,1)=p(1,0)=0$. This implies that the marginal distribution
for the control operation is \textbf{$p\left(k\right)=\sum_{l}p(k,l)=1/2$}
for $k=0,1$.\textbf{ }Accordingly, $\left\langle I^{(k,l)}\right\rangle =\sum_{k,l}p(k,l)\ln\frac{p(k,l)}{p(k)p(l)}=\textsf{H}_{Sh}(\frac{1}{2},\frac{1}{2})=\ln2$~nats
meaning maximum correlation between system and memory. In fact, the
measurement-based feedback protocol was designed to achieve this maximum.
Since the Shannon entropy, $\textsf{H}_{Sh}(p_{0}^{(0)},p_{1}^{(0)})$,
is upper bounded by $\ln2$~nats, we conclude that $\mathcal{I}_{gain}\leq\left\langle I^{(k,l)}\right\rangle $,
where the inequality is saturated in the limit $\beta\rightarrow0$
as can be noted in the experimental data displayed in Fig.~\ref{fig3}(b) of
the main text.

\textbf{Error analysis. }The main sources of error in the experiments
are small non-homogeneities of the transverse rf-field, non-idealities
in its time modulation, and non-idealities in the longitudinal field
gradient. In order to estimate the error propagation, we have used
a Monte Carlo method, to sampling deviations of the tomographic data
with a Gaussian distribution having widths determined by the variances
corresponding to such data. The standard deviation of the distribution
of values for the relevant information quantities is estimated from
this sampling. The variances of the tomographic data are obtained
preparing the same state ten times, taking the full state tomography
and comparing it with the theoretical expectation. These variances
include random and systematic errors in both state preparation and
data acquisition by QST. The error in each element of the density
matrix estimated from this analysis is about 1\%. All parameters in
the experimental implementation, such as pulses intensity and its
time duration, are optimized in order to minimize errors.

\end{document}